\documentclass[universe,review,submit,pdftex,moreauthors]{Definitions/mdpi}

\firstpage{1} 
\pubvolume{1}
\issuenum{1}
\articlenumber{0}
\pubyear{2025}
\copyrightyear{2025}
\datereceived{ } 
\daterevised{ } 
\dateaccepted{ } 
\datepublished{ } 
\hreflink{https://doi.org/} 

\usepackage{natbib,lineno}

\Title{The Impact of Dark Matter on Gravitational Wave Detection by Space-based Interferometers}

\TitleCitation{Dark Matter and Space-Based GW Detection}

\Author{Yuezhe Chen $^{1}$\orcidA{}, Pan-Pan Wang $^{2,}$*, Bo Wang $^{3,4}$, Rui Luo $^{1}$ and Cheng-Gang Shao $^{1}$}

\AuthorNames{Firstname Lastname, Firstname Lastname and Firstname Lastname}

\AuthorCitation{Chen, Y.; Wang, P.; Lastname, F.}

\address{%
$^{1}$ \quad National Gravitation Laboratory, MOE Key Laboratory of Fundamental Physical Quantities Measurement, and School of Physics, Huazhong University of Science and Technology, Wuhan 430074, People's Republic of China\\
$^{2}$ \quad College of Physics, Chongqing University, Chongqing 401331, China\\
$^{3}$ \quad Department of Astronomy, School of Physical Sciences, University of Science and Technology of China, 96 Jinzhai Road, Hefei, Anhui 230026, China\\
$^{4}$ \quad CAS Key Laboratory for Research in Galaxies and Cosmology, School of Astronomy and Space Science, University of Science and Technology of China, 96 Jinzhai Road, Hefei, Anhui 230026, China}

\corres{Correspondence: ppwang@hust.edu.cn}

\abstract{
The existence of dark matter is supported by multiple astrophysical observations, yet its particle nature remains unknown. The development of gravitational wave astronomy, especially with future space-based detectors such as LISA, provides new opportunities to study the interactions between dark matter and compact-object systems. This review summarizes the main dark matter candidates and their macroscopic distributions, and highlights three mechanisms through which dark matter can affect gravitational wave observations: (1) modifications to compact-object orbits and the dynamics of systems such as extreme mass-ratio inspirals, including dark matter spikes, dynamical friction, and potential perturbations; (2) gravitational lensing effects induced by the spatial distribution of dark matter, altering waveform amplitudes and phases; and (3) direct couplings between ultralight dark matter fields and detectors. As low-frequency gravitational wave detection techniques are proposed and continue to develop, these effects may offer a novel avenue for probing the properties of dark matter, and combining precise waveform modeling with multi-messenger observations could reveal insights into its microscopic structure.}


\begin{document}
\nolinenumbers
%
%

\section{Introduction}

Dark matter (DM) represents one of the major unresolved issues in contemporary physics and astronomy. The concept was first proposed by the astronomer Fritz Zwicky in 1933, who, through observations of galaxy cluster dynamics, found that the gravitational effects of these clusters could not be explained by visible matter alone, leading him to hypothesize the existence of an unseen form of matter, later termed DM~\cite{Intro1,Intro2}. In 1970, studies of galactic rotation curves by Vera Rubin and Kent Ford revealed that the rotational velocities of stars in the outskirts of galaxies were significantly higher than expected, further indicating the presence of a substantial amount of invisible matter in the universe~\cite{Intro3,Intro4}. Although DM has not yet been directly detected, its existence is widely accepted and constitutes a critical component of cosmology~\cite{Intro5}. DM plays a pivotal role in the evolution of the universe, the formation of galaxies, and the construction of large-scale structures~\cite{Intro5,Intro6}. Its study not only advances astrophysics but also provides modern physics with new insights into particle physics, gravity, and quantum mechanics~\cite{Intro2}.

Theoretical studies suggest that DM may consist of a variety of candidates, each with distinct production mechanisms in the early universe and potentially leaving detectable signals in experiments~\cite{Bertone2005, Feng2010}. Among these candidates, Weakly Interacting Massive Particles (WIMPs) have received particular attention~\cite{Jungman1996}. WIMPs are hypothetical fundamental particles that interact with the Standard Model (SM) only via weak interactions and commonly arise in several Standard Model extensions (SMEs)~\cite{Roszkowski2018, Arcadi2018}. These extensions can not only address the electroweak hierarchy problem but also naturally produce a relic abundance of DM consistent with observations via the thermal freeze-out mechanism~\cite{Kolb:1990vq}. If DM consists of WIMPs, they might be indirectly or directly detectable through scattering with ordinary matter, production in high-energy colliders, or annihilation in galaxies and more distant regions~\cite{Cushman2013, Aprile2018}. Besides WIMPs, other potential candidates include axions~\cite{Marsh2016}, primordial black holes~\cite{Carr_2022}, and supermassive non-thermal relic particles~\cite{Kuzmin1998}, each corresponding to distinct theoretical frameworks and experimental signatures. Despite ongoing efforts in modeling and observations, the true nature of DM remains elusive~\cite{Bertone2018}. Recently, gravitational wave (GW) has emerged as a novel observational tool, offering the potential to provide new insights into the nature of DM~\cite{Eda2013}.

The discovery of GWs has opened a new window for understanding the universe~\cite{Abbott2016_GW}. They allow for direct detection of compact objects, black hole mergers, supernova explosions, and other extreme physical processes~\cite{Maggiore2008, Sathyaprakash2009}, while also providing new avenues to investigate fundamental scientific questions, including the origin of the universe, DM, and dark energy~\cite{Kudoh2006}. Fundamentally, GWs are perturbations of spacetime produced by the asymmetric acceleration of massive objects~\cite{Einstein1916}. According to Einstein's field equations, when the quadrupole moment of a mass distribution changes asymmetrically, the system radiates energy in the form of GWs~\cite{Peters1963}. Typical sources of GWs include the inspiral and merger of binary neutron stars, binary black holes, and black hole–neutron star systems, non-axisymmetric collapse of massive stars, and quantum fluctuations in the early universe~\cite{Abbott2017_BNS, Abbott2021_catalog, Buonanno2007, Maggiore2000}. GWs exhibit a series of unique physical properties: they are transverse waves, with oscillations perpendicular to their propagation direction, manifesting as periodic stretching and compression of spacetime along their path; they propagate at the speed of light, reflecting their nature as spacetime perturbations~\cite{Misner1973}. Moreover, GWs possess strong penetration ability, being largely unaffected by matter absorption or scattering, and can directly carry pristine information about extreme cosmic events, traversing dense dust clouds and interstellar media~\cite{Thorne1987}. Consequently, GWs have become a crucial tool for probing deep astrophysical processes~\cite{Abbott2016_O3}. Analysis of GW waveforms and spectra allows inference of physical properties of astrophysical systems, including mass, spin, and orbital evolution~\cite{Blanchet2014}, while also enabling investigations into DM, dark energy, and testing general relativity in the strong-field regime~\cite{Yunes2013, Abbott2019_GRTests}. The rise of GW astronomy thus marks humanity’s entrance into an era of observational science based on “spacetime vibrations.”

In 2015, the LIGO Scientific Collaboration made the first direct detection of GWs from a binary black hole merger (GW150914)~\cite{Abbott2016_GW,2GW150914}, confirming the existence of GWs and inaugurating a new era of “GW astronomy.” Subsequently, ground-based interferometers such as LIGO, Virgo, and KAGRA have reported multiple detections, confirming GW events from various compact object systems, including binary neutron stars and binary black holes~\cite{Abbott2017_BNS, Abbott2021_catalog, Akutsu2021_KAGRA}. These observational achievements have significantly advanced our understanding of strong-field gravity, the formation and evolution of compact objects, and the origin of cosmic structures~\cite{Sathyaprakash2009, Maggiore2008}. However, ground-based detectors are limited by seismic noise and terrestrial disturbances, with sensitivity primarily covering the high-frequency band from ~10 Hz to 1 kHz, making them incapable of detecting lower-frequency GWs (mHz and below)~\cite{LIGO_Sensitivity, KAGRA_Sensitivity}. These low-frequency waves often originate from more massive and distant systems, such as supermassive black hole mergers, white dwarf binaries, or early-universe processes like phase transitions or cosmic string networks~\cite{Sesana2004, Sesana2005, Pau2017}. To extend the observational frequency range and access richer cosmological information, attention has turned to space-based GW detection. Space-based laser interferometers deploy multiple satellites over millions of kilometers to construct a stable interferometric system capable of high-precision measurements of low-frequency GWs~\cite{Amaro-Seoane2017, Hu2017}. Representative projects include the European-led LISA (Laser Interferometer Space Antenna), China’s Taiji Program, and Tianqin Project~\cite{Amaro-Seoane2017, Hu2017, Luo2016_Tianqin}. These space-based interferometers operate in the frequency range of 0.1 mHz–1 Hz, filling the observational gap between ground-based detectors and pulsar timing arrays, and providing key observational tools to study supermassive black hole mergers, extreme mass-ratio inspirals (EMRIs), early-universe phase transition backgrounds, and DM-induced gravitational effects~\cite{Babak2017, Barausse2014}.

 GWs can be influenced by the distribution of matter and energy in the universe~\cite{Takahashi2003}. As the dominant component of cosmic matter, DM may significantly affect GW signals~\cite{Eda2013, Macedo2013}. Theoretically, DM can influence both the generation and propagation of GWs through multiple mechanisms~\cite{Cardoso2019}. On one hand, the gravitational potential of DM can alter the propagation path and phase of GWs, leaving observable imprints on the waveform~\cite{Kavanagh2020}. On the other hand, DM can serve as an environmental background affecting the dynamical evolution of compact object systems~\cite{Barausse2014, Macedo2013}. For example, in EMRIs, DM halos can significantly modify orbital decay via dynamical friction~\cite{Eda2015}. Additionally, if DM is composed of scalar fields, its oscillatory behavior may induce characteristic perturbations in the phase and amplitude of GWs~\cite{Marsh2016, Brito2015}; in certain models, DM could even couple to GW detectors via scalar or tensor fields, leaving potentially detectable signals in future space-based observations~\cite{Baryakhtar2017, Hui2017}. Overall, GWs provide a novel observational avenue for probing DM, while DM, in turn, may leave measurable signatures on GW signals~\cite{Cardoso2019, Eda2015}.

This paper aims to provide a systematic review of the main effects of DM on space-based interferometers GW detection and related research progress. In Sec.~\ref{DM_can}, we briefly introduce the basic properties and common models of DM. In Sec.~\ref{DMGW}, we review the main effects of DM on GWs, where in Sec.~\ref{DM_sou}, we discuss the influence of DM on GW sources, including EMRIs and supermassive black hole binary systems, in Sec.~\ref{DM_pro}, we analyze modifications of GW propagation due to DM, in Sec.~\ref{DM_det}, we review the potential impact of DM on GW detectors. In Sec.~\ref{MultiMess}, we discuss multi-messenger strategies. Finally, in Sec.~\ref{sum}, we provide a perspective on the prospects of using space-based GW observations to constrain DM properties and advance cosmological studies. We hope this work provides a comprehensive view of the interplay between DM and GW detection.

\section{Dark Matter Candidates and Probing}\label{DM_can}
After decades of research, a diverse and extensive range of theoretical candidates for DM has emerged, spanning possibilities from fundamental particles to astrophysical-scale objects. Different models exhibit distinct characteristics in terms of mass range, interaction mechanisms, and cosmological origins, and their corresponding experimental signatures vary significantly. To systematically discuss the main research directions, this paper focuses on three representative candidate classes: WIMPs, ultralight DM (including axions, scalar field DM, vector DM, and dark photons), macroscopic DM halo models and self-interacting DM (SIDM). In the last section, we reviewed the current methods for detecting DM, including the use of space-based GW interferometers as the focus of this article for detection.

\subsection{Weakly Interacting Massive Particles}
WIMPs are hypothetical DM particles and currently represent one of the most extensively studied classes of DM candidates~\cite{Jungman1996}. They fall within the paradigm of “cold DM” due to their low velocities and typical masses in the GeV–TeV range, giving them non-relativistic characteristics. The defining feature of WIMPs is their coupling to SM particles via weak interactions—that is, they interact with ordinary matter only through the weak force (or interactions of comparable strength) and gravity~\cite{Bertone2005}. This interaction property allows WIMPs to be thermally produced in the early universe via interactions with SM particles and also makes them potentially detectable in laboratory experiments~\cite{Arcadi2018}.

From a theoretical perspective, the WIMP paradigm has two notable advantages. First, WIMPs can be generated through the thermal freeze-out mechanism in the early universe in a natural and simple manner, yielding a relic abundance consistent with current DM observations~\cite{Kolb:1990vq}. This typically occurs for particle masses near the weak scale, i.e., around the masses of $W^\pm$ and $Z_0$ bosons. Second, many theories that address the electroweak hierarchy problem through SMEs, such as supersymmetry (SUSY), naturally predict stable or long-lived particles at the weak scale, which can serve as WIMP candidates~\cite{Jungman1996}. These two reasons led to the so-called “WIMP miracle.”

During the thermal freeze-out process, WIMPs frequently interact with SM particles in the early universe through production, scattering, and annihilation, maintaining thermal equilibrium~\cite{Kolb:1990vq}. As the universe expands and cools, these interaction rates gradually decrease, and when they drop below the Hubble expansion rate, the WIMP number density effectively “freezes out,” resulting in the relic DM abundance observed today. Consequently, the parameter space of WIMPs (mass, scattering cross section, annihilation cross section) has become a major focus of both theoretical and experimental studies~\cite{Bertone2005}.

In many SMEs, WIMP DM candidates are introduced to protect the weak scale from quantum corrections~\cite{Jungman1996}. The most typical example is in SUSY, where, under R-parity conservation, the lightest supersymmetric particle (LSP) is stable; if this particle is electrically neutral and lacks strong interactions, it becomes a WIMP candidate~\cite{Martin1997}. In the Minimal Supersymmetric Standard Model (MSSM), WIMP candidates are typically linear combinations of the supersymmetric partners of the neutral Higgs bosons, $Z_0$, and photon (collectively termed “neutralinos”)~\cite{Ellis1984}. Neutralinos as WIMP candidates have been extensively studied and remain promising~\cite{Jungman1996}.

The “SM-connectable” nature of WIMPs provides multiple avenues for experimental searches. First, in direct detection experiments, WIMPs may scatter off atomic nuclei or electrons, producing measurable nuclear or electron recoils~\cite{Goodman1985}. Second, in regions with high DM density in the Milky Way or external galaxies, WIMPs may annihilate or decay into SM particles such as $\gamma$-rays, neutrinos, or cosmic rays, enabling indirect detection~\cite{Bertone2005}. Third, at high-energy colliders (e.g., the Large Hadron Collider, LHC), WIMP production may manifest as missing energy signatures~\cite{Abercrombie2020}. Fig.5 from \cite{PRL.131.041002} shows the 90\% confidence level upper limits on the spin-independent WIMP-nucleon cross section $\sigma_{SI}$ as a function of WIMP mass.

Recent advances in WIMP searches include the LUX-ZEPLIN (LZ) experiment, which achieved world-leading sensitivity using a multi-ton liquid xenon detector~\cite{Akerib2023}. No conclusive WIMP signals have been observed so far, but the upper limits on the WIMP-nucleon scattering cross section have been further tightened~\cite{Schumann2019}. A recent review notes: “Although the WIMP paradigm remains one of the most motivated DM frameworks, comprehensive searches using direct, indirect, and collider experiments have yet to yield definitive detection”~\cite{Abercrombie2020}. This indicates that, while the WIMP model is still under active investigation, the community is also considering its potential decline or the need to explore broader DM paradigms. Studies have highlighted that “WIMP searches have entered the ‘neutrino floor’ era, where experimental sensitivity approaches the limits set by neutrino backgrounds”~\cite{Billard2014}.

\subsection{Ultralight Dark Matter}
Ultralight DM (ULDM) refers to DM candidates with extremely low masses, typically in the range of $10^{-24}$ eV to $10^{-18}$ eV~\cite{Hu2000}. Unlike WIMP models with GeV–TeV mass scales, ULDM is generally realized as a bosonic field with an enormous occupation number, whose de Broglie wavelength can reach galactic or even larger macroscopic scales~\cite{Marsh2016}. In the early universe, such bosonic fields can be generated via the vacuum misalignment mechanism: the field initially deviates from the minimum of its potential during inflation or reheating and is nearly homogeneous over cosmological scales due to inflation~\cite{Preskill1983, Abbott1983}. As the universe expands and cools, when the Hubble parameter $H(t)$ drops to a scale comparable to the field mass $m$, the field begins coherent oscillations around the potential minimum. The energy density of the oscillating field evolves approximately as $\rho \sim a^{-3}$, behaving like cold DM and thus serving as a viable DM candidate~\cite{Sikivie2009}. Unlike thermal production mechanisms, ULDM is typically non-thermal and does not rely on thermal equilibrium with SM particles. This allows ULDM to more easily satisfy large-scale structure formation and other cosmological constraints. On galactic and sub-galactic scales, the wave-like properties of ULDM (such as quantum pressure and interference effects) can uniquely influence DM clustering, leading, for example, to solitonic cores at galactic centers and suppression of small-scale subhalo formation~\cite{Schive2014}.

\subsubsection{Axions}
The most widely studied realization of ultralight bosonic DM is the quantum chromodynamics (QCD)  axion~\cite{Peccei1977}. This novel scalar particle was originally proposed to solve the strong CP problem in QCD, a major naturalness issue in the SM~\cite{Kim1987, Dine1981}. In the SM Lagrangian, symmetry considerations allow the presence of a $\Theta$-term:
\begin{align}\label{axL}
	\mathcal{L} \subset -\Theta \frac{\alpha_s}{8\pi}G_{\mu\nu}\tilde{G}^{\mu\nu},
\end{align}
where $\Theta$ is a dimensionless parameter, $\alpha_s$ is the strong coupling constant, $G_{\mu\nu}$ is the gluon field strength tensor, and $\tilde{G}^{\mu\nu}$ is its dual. This term violates parity (P) and time-reversal (T) discrete symmetries, contributing proportionally to the neutron electric dipole moment~\cite{Crewther1979}. Consistency with experimental bounds requires $|\Theta| \le 10^{-10}$~\cite{Baker2006}. Since the natural range is $\Theta \in [-\pi, \pi)$, the strong CP problem asks why the effective $\Theta$ is so small.

To address this, Peccei and Quinn proposed a new global U(1) symmetry (PQ symmetry)~\cite{Peccei1977b}, which is anomalous under strong interactions and spontaneously broken at a high energy scale. At energies below the symmetry-breaking scale $f_a$, the theory predicts a small pseudo-Nambu–Goldstone field, the axion~\cite{Weinberg1978, Wilczek1978}. The axion couples to gluons similarly to Eq.~\eqref{axL}, with $\Theta \rightarrow \theta(x)=a(x)/f_a$. This coupling induces a potential for the axion field via QCD effects, minimizing $\theta(x)$ at zero and thereby solving the strong CP problem~\cite{DiVecchia1980}. The axion potential also gives rise to the axion mass~\cite{Sikivie2008}:
\begin{equation}
	m_a \simeq 6.0 \times 10^{-6} \ \text{eV} \left( \frac{10^{12} \ \text{GeV}}{f_a} \right).
\end{equation}

Axions can form DM through the ultralight boson mechanism described above~\cite{Preskill1983}. Notably, this mechanism is closely tied to the solution of the strong CP problem. Before the QCD phase transition in the early universe, the axion potential is nearly flat, allowing the field value $\theta = a/f_a \in [-\pi, \pi)$ to take any value. During the QCD transition at $T \sim 150$ MeV, a non-trivial potential develops, and the field is displaced from the new minimum, inducing vacuum misalignment~\cite{Turner1986}. The resulting axion field oscillations act as DM, with relic density~\cite{Wantz2010}:
\begin{equation}\label{axde}
	\Omega_a h^2 =0.12 \left(\frac{f_a}{9\times 10^{11} \ \text{GeV} }\right)^{1.165}\left[\bar{\theta}^2_i + (\delta \theta_i)^2\right]F,
\end{equation}
where $\bar{\theta}_i = a(t_i)/f_a$ is the average initial misalignment angle over the Hubble volume, $\delta \theta_i$ is the RMS fluctuation, and $F$ accounts for anharmonic corrections, typically $\mathcal{O}(1)$ for $f_a \le M_{\text{Pl}}/10$, with $M_{\text{Pl}}$ the Planck mass.

The breaking of PQ U(1) symmetry can interplay with cosmic inflation~\cite{Lyth1992}. If the PQ symmetry breaks after inflation, the observable universe comprises many causally disconnected patches, each with independent initial field values $\theta_i$. The axion relic density is then proportional to the mean-square of these angles, corresponding to the fluctuation term in Eq.~\eqref{axde}, with $(\delta\theta_i)^2 \sim 1$. Topological defects such as domain walls and cosmic strings can also form, providing additional axion sources and other effects~\cite{Davis1986}.

If PQ symmetry breaks before inflation, a single $\theta_i = \bar{\theta}_i \gg |\delta\theta_i|$ region can encompass the observable universe~\cite{Tegmark2006}. This allows for correct relic density with smaller $\theta_i$ and correspondingly larger $f_a$. However, axion DM models in which the Peccei--Quinn symmetry is broken before inflation are subject to stringent constraints from primordial isocurvature perturbations. During inflation, quantum fluctuations of the axion field generate fluctuations in the initial misalignment angle, with variance
\begin{equation}
	\delta \theta_i \simeq \frac{H_{\rm inf}}{2\pi f_a},
\end{equation}
where $H_{\rm inf}$ is the Hubble parameter during inflation. These fluctuations lead to cold DM isocurvature perturbations, whose power spectrum can be expressed as
\begin{equation}
	\mathcal{P}_{\rm iso} \simeq 
	\left(\frac{\partial \ln \rho_a}{\partial \theta_i}\right)^2
	\left(\frac{H_{\rm inf}}{2\pi f_a}\right)^2
	\simeq
	\left(\frac{H_{\rm inf}}{\pi f_a \bar{\theta}_i}\right)^2 ,
\end{equation}
where $\rho_a$ denotes the axion energy density and vacuum misalignment is assumed to dominate the relic abundance. The resulting isocurvature contribution to the total primordial perturbations is tightly constrained by observations of the cosmic microwave background. In particular, Planck measurements impose an upper bound on the isocurvature fraction $\beta_{\rm iso} \equiv \mathcal{P}_{\rm iso}/(\mathcal{P}_{\rm ad}+\mathcal{P}_{\rm iso})$ at the percent level~\cite{Planck2018}.
These bounds translate into an upper limit on the inflationary energy scale $H_{\rm inf}$ as a function of the axion decay constant $f_a$ and the initial misalignment angle $\bar{\theta}_i$, significantly restricting pre-inflationary axion DM models, especially for large $f_a$. By contrast, in post-inflationary scenarios, the axion field takes random values in different causally disconnected regions, and the resulting isocurvature perturbations are effectively averaged out, rendering CMB isocurvature constraints much less severe, although uncertainties associated with topological defect contributions remain~\cite{Kawasaki2015}. Assuming vacuum misalignment dominates the axion DM relic density, the QCD axion mass is typically expected to lie in the range $m_a = 10^{-12}\ \text{eV} - 10^{-4}\ \text{eV}$, corresponding to $f_a \sim 10^{12} \ \text{GeV} - M_{\text{Pl}}$~\cite{Kim2010}. Recent reanalyses of precision laboratory experiments have led to new constraints on ultralight axion-like DM; in particular, significantly strengthens laboratory limits on axion couplings in the $10^{-24}-10^{-21}$eV mass window, partially surpassing existing astrophysical bounds~\cite{axion2025}.

Beyond QCD axions, there exists a broader class of axion-like particles (ALPs), which also arise from spontaneous breaking of continuous symmetries and are pseudo-Nambu–Goldstone bosons, but with masses and couplings unconstrained by QCD~\cite{Jaeckel2010}. ALPs appear naturally in various high-energy theories, such as string compactifications with moduli fields~\cite{Svrcek2006}. Like QCD axions, ALPs can achieve relic abundance through the misalignment mechanism, behaving as cold DM on cosmological scales~\cite{Marsh2016}. With masses spanning $10^{-22}$ eV to eV or higher, ALPs exhibit diverse physical effects: ultralight ALPs induce wave-like phenomena and central density flattening at galactic scales, while heavier ALPs behave more like conventional particle DM~\cite{Hui2017}.

\subsubsection{Scalar Field Dark Matter}
Scalar field DM (SFDM) consists of spin-0 bosons with typical masses in the range $10^{-22}\text{eV}-10^{-10}\text{eV}$~\cite{Hui2017}. Axions are formally a subset of SFDM but are treated separately due to their unique theoretical origin and importance~\cite{Marsh2016}. Due to their extremely low mass, the de Broglie wavelength
\begin{equation}
	\lambda_{\text{dB}}=\frac{h}{m_\phi v},
\end{equation}
with $h$ the Planck constant and $v$ the particle velocity, can extend to galactic or even cluster scales, leading to significant quantum coherence and wave phenomena at astrophysical scales~\cite{Hu2000}. These properties impart SFDM with behavior distinct from classical cold DM (CDM) in galaxy formation and dynamics.

SFDM typically arises from vacuum misalignment or phase transitions in the early universe~\cite{Sikivie2009}. After inflation or reheating, the scalar field may be displaced from the minimum of its potential $V(\phi)$, evolving as quasi-harmonic oscillations during cosmic expansion. When $H \sim m_\phi$ (i.e., $3H \sim m_\phi$), the field begins oscillating about the minimum, with mean energy density scaling as $\rho_\phi \propto a^{-3}$, behaving as non-relativistic matter. Quantum pressure suppresses small-scale density peaks, alleviating the “core–cusp” problem of classical CDM~\cite{Schive2014}.

In the non-relativistic limit, SFDM can be expressed via the Madelung representation~\cite{Chavanis2011}:
\begin{align}
	\phi(\vec{x},t)=\frac{1}{\sqrt{2m_\phi}}\left[\Psi(\vec{x},t)e^{-im_\phi t}+\Psi^* (\vec{x},t)e^{im_\phi t}\right],
\end{align}
with evolution governed by the Schrödinger–Poisson system:
\begin{align}
	i\hbar \frac{\partial \Psi}{\partial t}=\frac{\hbar^2}{2m_\phi}\nabla^2 \Psi+m_\phi \Phi \Psi,\qquad \nabla^2 \Phi = 4\pi G |\Psi|^2.
\end{align}
This system demonstrates the quantum fluid nature of SFDM, with quantum pressure preventing over-collapse at halo centers, producing flat cores consistent with observed rotation curves~\cite{Hui2017}. In gravitationally bound systems, SFDM can also form quasi-stable Bose–Einstein condensates (BECs) or boson stars, reinforcing its relevance in astrophysical modeling~\cite{Liebling2012}.

The mass of SFDM determines its cosmological effects. For $m_\phi \sim 10^{-22}$ eV, wave effects are maximized, forming fuzzy DM (FDM) with interference patterns and quasi-periodic density distributions at galactic scales~\cite{Schive2014}. For $m_\phi \ge 10^{-18}$ eV, wave effects diminish and dynamics approach classical CDM behavior. Different mass ranges therefore yield distinguishable consequences for structure formation, galactic dynamics, and observational signatures~\cite{Marsh2016}.

SFDM wave-like features may also produce observable effects through various mechanisms. In strong gravity environments, superradiant clouds can form around black holes or neutron stars, modifying spin distributions and GW emissions~\cite{Arvanitaki2015}. At galactic scales, quantum interference may leave signals in lensing or luminosity distributions~\cite{Schive2014}. At cosmological scales, SFDM evolution can imprint signatures on the cosmic microwave background (CMB) temperature and polarization power spectra~\cite{Hlozek2015}.

\subsubsection{Vector Field Dark Matter}
Vector field DM (VFDM) is a hypothetical candidate composed of spin-1 bosons, in contrast to the spin-0 nature of scalar DM~\cite{Nelson2011}. VFDM may arise from gauge fields or massless particles (e.g., photons), acquiring mass through symmetry-breaking mechanisms (Higgs or Stueckelberg) and coupling to gravity, providing a potential pathway to understanding DM~\cite{Graham2016}.

VFDM is inherently directional due to the vector field structure $A_\mu$, with dynamics described by classical or quantized Yang–Mills fields coupled to gravity. The Lagrangian is typically~\cite{Nelson2011}:
\begin{align}
	\mathcal{L}=-\frac{1}{4}F_{\mu\nu}F^{\mu\nu}+\frac{1}{2}m^2A_\mu A^\mu,
\end{align}
where $F_{\mu\nu}=\partial_\mu A_\nu -\partial_\nu A_\mu$ and $m$ is the vector particle mass. The energy density evolves like cold DM as the universe expands.

At cosmological scales, the coherence length of VFDM is inversely related to mass. Ultralight masses can yield de Broglie wavelengths comparable to galactic scales, producing wave-like phenomena similar to FDM~\cite{Nelson2011}. The directional nature may induce anisotropic gravitational effects, affecting halo shapes and galactic dynamics~\cite{Marsh2016}. Coupling to other fields, such as the electromagnetic or scalar fields, can generate observable signals including low-frequency GW backgrounds, CMB polarization distortions, or dark radiation~\cite{Graham2016}. Around compact objects, VFDM may influence black hole or neutron star spin evolution, accretion disk dynamics, and GW emission~\cite{Arvanitaki2015}.

\emph{Dark photons} (also denoted $A'$ or $\text{U(1)}_\text{D}$ photons) are a class of light vector bosons associated with an additional U(1) gauge symmetry in DM models~\cite{Essig2013}. Analogous to SM photons, dark photons can kinetically mix with the electromagnetic field, producing potentially observable signals in laboratory or astrophysical settings~\cite{Redondo2009}. Dark photons may constitute DM themselves or mediate interactions in more complex DM models. In the ultralight regime, dark photons can form coherent fields on cosmological scales, generating quasi-stable wave-like structures~\cite{Nelson2011}; under high-energy astrophysical or laboratory conditions, they may be indirectly detected via decay, resonant absorption, or missing energy signatures~\cite{Essig2013}.

\subsection{Dark Matter Halos — NFW, Burkert, and Hernquist Models}
DM halo models are fundamental theoretical tools for describing the spatial distribution of DM in galaxies and galaxy clusters~\cite{Mo2010}. Different models correspond to distinct galactic structural characteristics and formation mechanisms. Among them, the Navarro–Frenk–White (NFW) model, Burkert model, and Hernquist model are the most commonly used empirical density profiles, providing a theoretical basis for understanding DM clustering across different scales~\cite{Navarro1996,Burkert1995,Hernquist1990}.

The NFW model originates from cold DM cosmological numerical simulations and is the most widely adopted empirical profile for galactic and cluster-scale DM halos~\cite{Navarro1996}. Its density distribution is given by
\begin{align}
	\rho(r)=\frac{\rho_0}{\frac{r}{r_s}\left(1+\frac{r}{r_s}\right)^2},
\end{align}
where $r$ is the distance from the halo center, $r_s$ is a scale radius marking the transition region of the density profile, and $\rho_0$ is the central density. In the small-radius limit, $\rho(r) \propto r^{-1}$, while at large radii, $\rho(r) \propto r^{-3}$. This “double power-law” structure reveals the cuspy nature of DM in the central regions and the rapid decline in the outskirts~\cite{Navarro1996}. The NFW profile has been extensively validated through numerical simulations and gravitational lensing analyses of galaxy clusters~\cite{Mo2010}.

The Burkert model is an empirical “cored” density profile, mainly used to describe DM distributions in low surface brightness and dwarf galaxies~\cite{Burkert1995}. Its density function is
\begin{align}
	\rho(r)=\frac{\rho_0}{\left(1+\frac{r}{r_s}\right)\left(1+\frac{r^2}{r_s^2}\right)},
\end{align}
which approaches a constant-density core as $r \rightarrow 0$, while decaying as $\rho(r) \propto r^{-3}$ for $r \gg r_s$. The Burkert profile successfully accounts for the flat-core structures observed in the rotation curves of dwarf galaxies, in contrast to the cuspy distribution predicted by the NFW model, and is often considered a phenomenological correction or supplement to CDM simulation results~\cite{Burkert1995}.

The Hernquist model was originally proposed to describe the luminosity and mass distributions of elliptical galaxies~\cite{Hernquist1990}, with a density profile of
\begin{align}
	\rho(r)=\frac{\rho_0}{\left(\frac{r}{r_s}\right)\left(1+\frac{r}{r_s}\right)^3},
\end{align}
where the central density scales as $r^{-1}$, while at large radii it decreases more steeply as $\rho(r) \propto r^{-4}$. Compared with the NFW profile, the Hernquist distribution has a sharper outer decline, making it suitable for describing highly concentrated elliptical galaxies or spheroidal halo structures~\cite{Hernquist1990}.

Overall, the NFW model is most appropriate for cold DM halos of massive galaxies and clusters; the Burkert model performs better for low-mass, low-luminosity galaxies, capturing the flat core density; and the Hernquist model excels in describing concentrated mass distributions in elliptical and spheroidal systems. Together, these three profiles complement each other across different mass scales, forming the main theoretical framework for current studies of DM halo structures~\cite{Mo2010}.
\subsection{Self-Interacting Dark Matter}

SIDM constitutes an important extension of the standard CDM paradigm. Its basic assumption is that, in addition to gravitational interactions, DM particles experience non-gravitational self-interactions~\cite{Intro6}. These interactions are typically modeled as elastic scattering processes between DM particles, and their strength is commonly characterized by the scattering cross section per unit mass, $\sigma/m_{\chi}$, where $\sigma$ denotes the DM--DM scattering cross section and $m_{\chi}$ is the DM particle mass. Phenomenologically, values in the range
\begin{equation}
	\frac{\sigma}{m_{\chi}}\sim0.1-10 \,\mathrm{cm}^2\,\mathrm{g}^{-1},
\end{equation}
are known to produce significant effects on galactic scales, while remaining consistent with constraints from large-scale structure formation and CMB observations.

It is important to emphasize that SIDM does not correspond to a specific particle DM candidate. Rather, it represents an effective framework that parameterizes the microscopic properties of DM. As discussed in the previous subsection, in the conventional CDM model DM is treated as collisionless, leading numerical simulations to predict halos with steep central density cusps. However, kinematic observations of dwarf galaxies and low surface brightness galaxies often favor nearly constant-density cores. By introducing self-interactions, SIDM allows DM particles to undergo multiple scatterings within galactic halos, resulting in a redistribution of energy and momentum that naturally produces cored density profiles.

At the macroscopic level, the dynamical evolution of SIDM can be described by a Vlasov--Poisson equation supplemented by a collision term,
\begin{equation}
	\frac{\partial f}{\partial t}+\mathbf{v}\cdot\nabla_{\mathbf{x}}f-\nabla_{\mathbf{x}}\Phi\cdot\nabla_{\mathbf{v}}f=C[f],
\end{equation}
where $f(\mathbf{x},\mathbf{v},t)$ denotes the phase-space distribution function of DM, $\Phi$ is the gravitational potential, and $C[f]$ represents the collision term induced by DM self-interactions. When the scattering rate is sufficiently low, the collision term only becomes important within galactic halos, while remaining negligible on larger cosmological scales, thereby preserving the success of CDM in describing large-scale structure.

Many contemporary SIDM models further generalize this framework by introducing velocity-dependent self-interactions, $\sigma=\sigma(v)$, where $v$ is the typical relative velocity of DM particles in a halo~\cite{Vogelsberger2012,Kaplinghat2015}. Such models naturally yield strong self-interactions on dwarf-galaxy scales, where velocities are low, while significantly suppressing interactions on galaxy-cluster scales, thus simultaneously satisfying small-scale structure observations and upper limits from systems such as the Bullet Cluster.

From an observational perspective, SIDM models are constrained by a combination of astrophysical and cosmological probes. Colliding galaxy clusters typically impose an upper bound of
\begin{equation}
	\frac{\sigma}{m_{\chi}}\lesssim \mathcal{O}(1)\,\mathrm{cm}^2\,\mathrm{g}^{-1},
\end{equation}
while galaxy rotation curves, density profiles of dwarf galaxies, and the ellipticity of galaxy clusters provide complementary constraints on the SIDM parameter space~\cite{Robertson2016}. Overall, SIDM offers a theoretically well-motivated and observationally viable framework that preserves the successes of CDM on large scales while providing a natural explanation for DM distribution features on galactic scales.

\textit{Mirror matter.} Beyond the phenomenological framework of SIDM, mirror matter represents a theoretically well-motivated class of dark-sector models in which the DM consists of a mirror copy of the SM related to the visible sector by a discrete parity symmetry~\cite{Blinnikov1983,Khlopov1991}. Interactions between ordinary and mirror particles are predominantly gravitational, while possible non-gravitational portals are tightly constrained~\cite{Lee1956,Foot1991}. Within the mirror sector, mirror electromagnetic and nuclear interactions lead naturally to non-negligible self-interactions and, in some realizations, dissipative dynamics. As a result, mirror matter can exhibit rich astrophysical behavior, potentially affecting the formation and internal structure of DM halos and compact-object environments~\cite{Foot2014}. In this sense, mirror matter can be viewed as a concrete microscopic realization underlying a broader class of self-interacting or dissipative DM scenarios, while remaining consistent with current cosmological and astrophysical constraints.

Besides the DM candidates mainly discussed in this review, a wide range of alternative models have been proposed in which the dark sector possesses a richer internal structure and a non-standard cosmological evolution. For example, cosmological hydrodynamical simulations indicate that a subdominant component of dissipative atomic DM can cool efficiently and form macroscopic structures such as dark disks and compact clumps, potentially affecting galactic morphology and baryonic evolution even when its fractional abundance is small~\cite{Roy2023}. Other studies investigate dark-sector phase transitions occurring after primordial nucleosynthesis, sometimes referred to as a “dark big bang,” which can account for the observed DM abundance and may generate stochastic GW  backgrounds accessible to future observations~\cite{Freese2023}. Additional frameworks consider the existence of multiple or mirrored copies of the Standard Model, where hidden-sector baryons or string- and brane-based constructions provide viable DM candidates with particle spectra related to that of visible matter~\cite{Guendelman2025,Dvali2009}. In strongly interacting dark sectors, gravitational collapse can also lead to the formation of compact objects such as dark baryon–induced black holes or long-lived black hole relics~\cite{Profumo2025}. Given the review nature of the present work, a detailed discussion of these scenarios is beyond its scope, and we refer interested readers to the original literature for further details.

\subsection{Probes of Dark Matter}
A wide variety of observational and experimental approaches have been developed to probe the nature of DM, each targeting different aspects of its microphysical properties or macroscopic behavior. At this stage, it is useful to briefly summarize these traditional probes and clarify the regimes in which they are most effective, before discussing alternative and complementary approaches in later sections.

Direct detection experiments search for non-gravitational interactions between DM particles and ordinary matter, typically via nuclear recoils induced by elastic scattering~\cite{Feng2010}. These experiments place stringent bounds on the interaction cross section for particle DM in the GeV–TeV mass range, but their sensitivity rapidly decreases for very light DM, ultra-heavy candidates, or scenarios in which DM couples extremely weakly—or not at all—to Standard Model particles.

Indirect detection searches aim to identify annihilation or decay products of DM, such as gamma rays, charged cosmic rays, or neutrinos~\cite{Bertone2005}. These observations constrain annihilation cross sections or decay rates under specific assumptions about the DM distribution and astrophysical backgrounds. As a result, they are primarily sensitive to particle DM models with sizable self-annihilation or decay channels and are ineffective for stable, non-annihilating, or purely gravitational DM candidates.

Cosmological and large-scale structure observations, including galaxy surveys, cosmic microwave background measurements, and weak gravitational lensing, probe the clustering properties of DM on large scales~\cite{Bullock2017}. These methods tightly constrain deviations from the standard cold and collisionless DM paradigm, such as strong self-interactions or significant free-streaming effects. However, they are largely insensitive to the detailed dynamics of DM in strongly gravitating, highly nonlinear environments.

Taken together, these traditional probes have established a broad and increasingly precise picture of DM across particle, galactic, and cosmological scales. At the same time, they leave open important regions of parameter space—particularly for DM candidates whose interactions are predominantly gravitational, collective, or environmental in nature, and for scenarios in which DM effects are most pronounced in compact or high-density astrophysical systems. In this context, space-based GW  interferometers provide a qualitatively new observational window. By probing the dynamics of compact objects, the propagation of GWs, and the properties of strong-field environments, GW  observations offer sensitivity to aspects of DM that are difficult or impossible to access with traditional methods. The following sections will focus on how GWs, and in particular observations with space-based detectors, can be used to probe DM models in a complementary and, in some cases, unique manner.

\section{Dark Matter Effects on Gravitational Waves}\label{DMGW}
\begin{figure}[htbp]
	\includegraphics[width=1\textwidth]{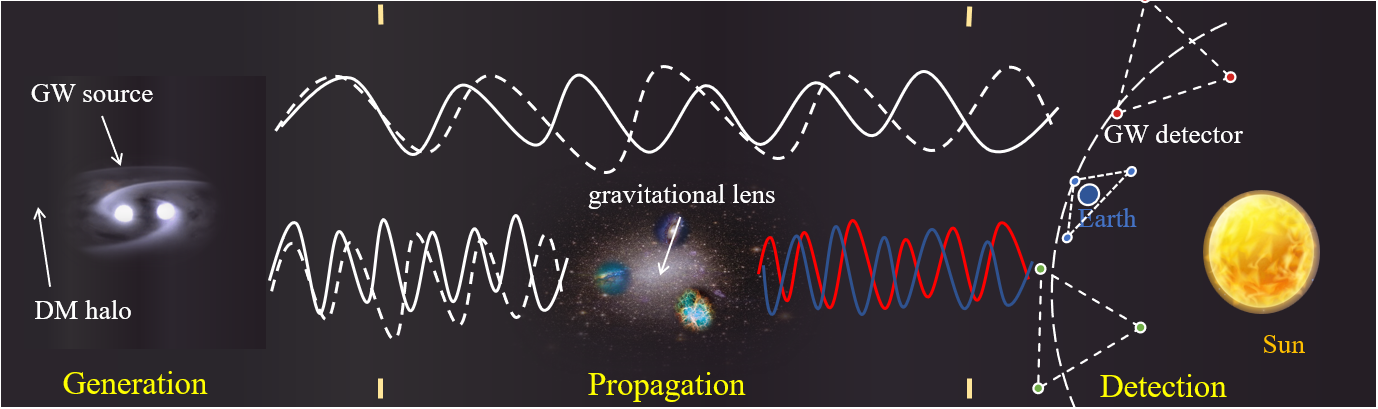}
	\caption{\label{DM_GW}Schematic illustration of the effects of DM on GWs from their generation to detection.
		The white solid and dashed curves represent GWs emitted by the source with and without the influence of DM, respectively.
		The blue and red solid curves denote the corresponding GW signals after propagation through the Universe, where they may be further modified by DM-induced gravitational lensing.
		This figure is intended as a conceptual illustration of the different stages at which DM can affect GW observations, rather than a quantitative prediction.
		The impact of DM on GW detectors is not explicitly shown.
		The three space-based detectors orbiting the Sun, from top to bottom, are LISA, TianQin, and Taiji.}
\end{figure}

In this section, we review the main effects of DM on GWs. As schematically illustrated in Figure~\ref{DM_GW}, the evolution of a GW signal from its production to its detection can be divided into three stages: the generation of GWs at the source, their propagation through the Universe, and their detection by GW observatories. Depending on the specific DM models and the associated physical effects, DM may affect each of these stages. The corresponding impacts will be discussed in detail in the following subsections.

\subsection{Effects of Dark Matter on Gravitational Wave Sources}\label{DM_sou}
We now discuss the impact of DM on multiple GW sources, including EMRIs, supermassive black hole binaries, and compact binary systems, which fall within the frequency band of space-based interferometers. DM primarily affects these sources by modifying the gravitational potential of the system or altering the surrounding environment. The specific mechanisms of these influences are detailed below.

\subsubsection{Effects of Dark Matter on EMRIs}
With the rapid development of GW astronomy, EMRIs have emerged as key sources for future space-based detectors due to their long-lasting, coherent, and parameter-rich millihertz GW signals~\cite{Amaro-Seoane2017, Babak2017}. EMRIs consist of a low-mass compact object (e.g., a stellar-mass black hole, neutron star, or white dwarf) slowly inspiraling around a supermassive black hole (SMBH). Their waveforms contain abundant information about the environment surrounding the central black hole, making them natural probes for studying DM distributions in the vicinity of black holes~\cite{Macedo2013}.

In recent years, the distribution of DM in the vicinity of compact objects, particularly around intermediate-mass and supermassive black holes, has been suggested to form highly concentrated spike structures (DM spikes)~\cite{Gondolo1999}. Such spikes were first proposed by Gondolo and Silk, who noted that if the central black hole grows adiabatically and slowly, the surrounding pre-existing DM halo could be compressed into a steep power-law profile. Subsequent studies, however, indicate that in realistic strong-gravity environments, traditional power-law approximations are insufficient to accurately describe the steady-state behavior of DM near the event horizon~\cite{Sadeghian2013}.

Sadeghian, Ferrara, and Will (SFW) developed a more self-consistent phase-space approach within the framework of general relativity to describe the steady-state distribution of DM around a Schwarzschild black hole~\cite{Sadeghian2013}. They assumed DM as a cold, collisionless particle system, described by a distribution function $f(E,L)$ in energy-angular momentum space. By directly solving for the allowed bound orbits in a GR spacetime and integrating over 4-momentum, they obtained the mass density at any radius. SFW computed the DM spike generated by adiabatic growth of a Hernquist halo near a Schwarzschild black hole. This method does not require assuming a “power-law” density a priori; rather, the inner saturation and outer power-law structure naturally emerge from orbital dynamics, revealing relativistic corrections to DM spikes in strong gravitational fields~\cite{Sadeghian2013}.

The self-consistent SFW framework also provides a theoretical foundation for assessing the effects of DM on GW signals, annihilation radiation, and EMRI dynamics~\cite{Sadeghian2013}. DM spikes formed under this framework not only significantly enhance the density near black holes but also modify the orbital distribution of particles in the strong-field region, resulting in additional perturbations to compact object orbits. EMRIs, as long-term inspiraling systems of low-mass compact objects around SMBHs, are highly sensitive to minute environmental perturbations, making them ideal tools to probe the properties of DM spikes~\cite{Eda2013}. Embedding an EMRI system in a DM-spike background alters the orbital decay rate, precession behavior, energy dissipation, and final inspiral trajectory.

Once the formation mechanisms are established, the effects of DM spikes and broader DM environments on EMRIs have become a major focus of GW theory~\cite{Li2022}. EMRIs involve stellar-mass compact objects spiraling around SMBHs for tens of thousands to millions of orbital cycles, making them highly sensitive to tiny dynamical perturbations. The primary mechanisms through which DM affects EMRIs are: modifications to spacetime geometry, orbital energy and angular momentum loss due to dynamical friction, and mass/spin evolution induced by DM accretion~\cite{Barausse2015}.

In the absence of DM, energy and angular momentum changes of EMRIs are mainly driven by GW radiation:
\begin{align}
	\left(\frac{dE}{dt}\right)_{\text{GW}}=g_{E}(m,M,p,e),\qquad \left(\frac{dL_z}{dt}\right)_{\text{GW}}=g_{L_z}(m,M,p,e),
\end{align}
where $m$ represents the mass of secondary, $M$ represents the mass of the centeral BH, $p$ and $e$ respectively represent the semi-latus rectum and eccentricity of the track. First, high-density DM spikes directly modify the gravitational potential near the black hole, changing the effective spacetime background for the EMRI secondary. This effect can be viewed as an additional mass perturbation on the Kerr (or Schwarzschild) background, modifying orbital precession, eccentricity evolution, and semimajor axis shrinkage rates~\cite{Eda2013}. Several studies have shown that, under typical spike distributions, cumulative orbital phase shifts may reach levels detectable by space interferometers. This effect is particularly pronounced in the relativistic steady-state distribution framework of SFW, which predicts a relativistic “saturation zone” near the event horizon, contributing systematically different orbital dynamics than standard Newtonian power-law models.

Second, DM distributions induce dynamical friction, affecting EMRI evolution. As the secondary moves through the DM medium, it generates a gravitational wake, resulting in a retarding force that causes additional orbital energy and angular momentum loss~\cite{Macedo2013}. This effect is more significant in high-density or low-velocity orbital phases and can be generalized to a relativistic context via the Chandrasekhar dynamical friction formula:
\begin{align}
	\left(\frac{dE}{dt}\right)_{\text{DF}}&=F_{\text{DF}}\cdot v=-\frac{4\pi \rho(r,<v) G^2 m^2 \gamma^2 \left[1+\left(\frac{v}{c}\right)^2\right]^2}{v}\ln\Lambda,\\
	\left(\frac{dL_z}{dt}\right)_{\text{DF}}&=r\cdot F_{\text{DF}} \frac{r \dot{\Psi}}{v}=-\frac{4\pi \rho(r,<v) G^2 m^2 r^2 \dot{\Psi}\gamma^2 \left[1+\left(\frac{v}{c}\right)^2\right]^2}{v^2}\ln\Lambda,
\end{align}
where $F$ represents dynamic friction, $v$ represents the velocity of secondary, $\rho(r,<v)$ represents the density of DM particles at radius $r$ whose velocity is less than the velocity of secondary, $\gamma$ is the Lorentz factor, and $\ln\Lambda$ is the Coulomb logarithm. Dynamical friction generally accelerates orbital decay, shortening the time to merger and causing additional phase shifts in the GW signal~\cite{Becker2022}.

Third, DM accretion can affect the intrinsic properties of the secondary. If DM particles are captured by the compact object, accretion modifies its mass and spin over time, altering orbital parameters and instantaneous GW frequency~\cite{Nichols2023}. Although generally weaker than the previous two mechanisms, this effect can be enhanced in certain scenarios (e.g., SIDM or compact DM cores), and thus must be considered in high-precision waveform modeling. Consequently, the total evolution of energy and angular momentum in an EMRI system immersed in DM is:
\begin{align}
	\frac{dE}{dt}=\left(\frac{dE}{dt}\right)_{\text{GW}}+\left(\frac{dE}{dt}\right)_{\text{DF}}+\left(\frac{dE}{dt}\right)_{\text{acc}},\\
	\frac{dL_z}{dt}=\left(\frac{dL_z}{dt}\right)_{\text{GW}}+\left(\frac{dL_z}{dt}\right)_{\text{DF}}+\left(\frac{dL_z}{dt}\right)_{\text{acc}}.
\end{align}
Specifically, in high-density regions, EMRI secondaries may experience faster energy loss and orbital decay, shortening merger timescales and inducing subtle modifications in GW spectra and waveforms.

\begin{figure}[htbp]
	\includegraphics[width=1\textwidth]{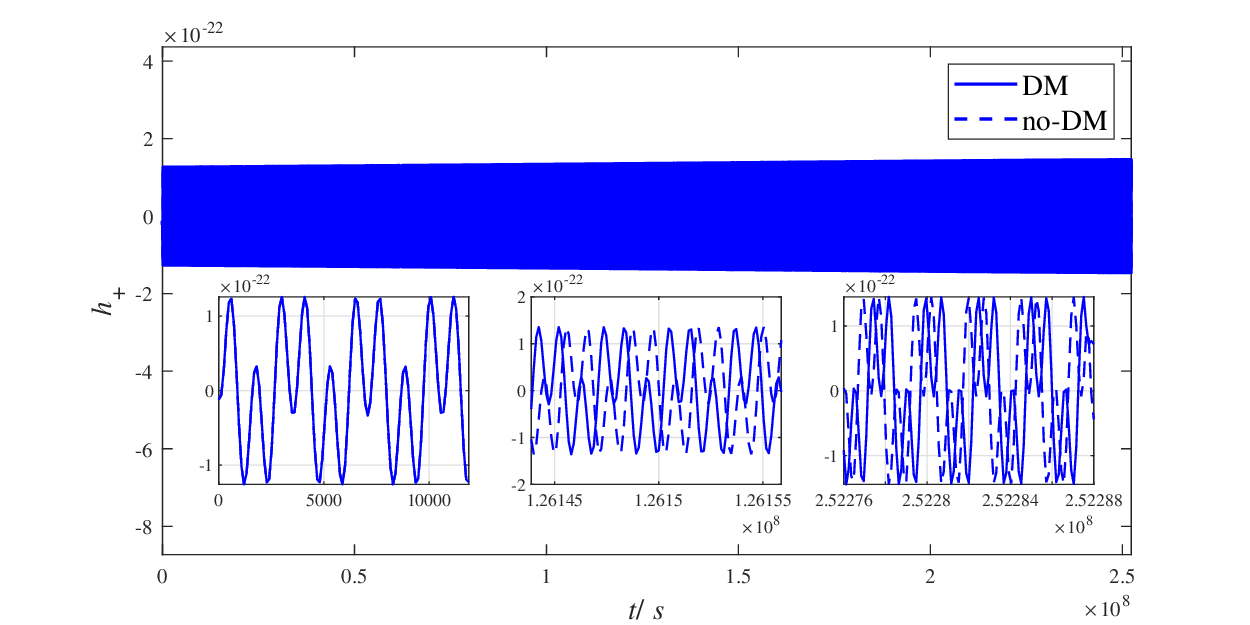}
	\caption{\label{EMRIs2} EMRI waveforms under constant DM models. Solid (dashed) lines show GW waveforms with (without) dynamical friction effects included.The main panel displays the waveform comparison over the entire inspiral evolution, which spans a very long timescale characteristic of EMRIs.
	The inset windows show zoomed-in waveform segments at the early, intermediate, and late stages of the inspiral, respectively, highlighting the local phase evolution and the cumulative impact of dynamical friction.}
\end{figure}

Theoretical studies by X.J. Yue et al. indicate that DM micro-spikes around black holes could enhance the event rate of IMRIs/EMRIs by several orders of magnitude~\cite{APJ.874.34}. Using NFW, Burkert, and Hernquist DM profiles, the N. Dai et al~\cite{PRD.110.084080,MN10.1093} and Z.C. Zhang et al~\cite{PRD.110.103008} conducted detailed modeling of DM halos around SMBHs and coupled these with EMRI orbital dynamics. Results show that DM halos can significantly alter EMRI trajectories and GW phase evolution, as illustrated in Fig.1 from ~\cite{PRD.110.084080} and Figure~\ref{EMRIs2} in this papper, suggesting that such effects may be detectable by future high-precision GW detectors. More recent work further indicates that DM density fluctuations (ULDM oscillations) may induce noise-like structures in EMRI waveforms, and scalar or vector ULDM fields could resonate with EMRI orbits, producing phase jumps or transient frequency drifts~\cite{Brax2024}.

In summary, DM influences EMRI orbital evolution and GW emission through modifications of spacetime, dynamical friction, and accretion effects. Given the extremely high phase accumulation of EMRI signals, future space-based detectors (such as LISA, TianQin, and Taiji) will have the capability to probe these subtle perturbations. By comparing waveforms in DM-present and vacuum backgrounds through template matching and parameter estimation, it may be possible to infer the DM distribution around SMBHs, thereby providing new observational constraints on DM spike formation, DM-strong gravity interactions, and the evolutionary history of galactic centers.

\subsubsection{Effects of Dark Matter on Compact Binary Systems}
Compact binary systems, such as binary black holes or binary neutron stars, due to their strong gravitational fields and rapid orbital motion, are crucial laboratories for studying extreme gravitational physics and GW astrophysics~\cite{Abbott2016}. During their long-term evolution, these systems lose orbital energy and angular momentum through GW emission, eventually entering the late inspiral phase and finally merging~\cite{Peters1964}. As possible interactions between DM particles and compact objects become a forefront topic in both theoretical and observational research, the impact of DM on orbital evolution, internal structure, and GW signals of compact binaries has attracted increasing attention~\cite{Barausse2014}.

DM accretion is an important mechanism influencing binary orbital evolution. When a compact binary resides in a galactic DM halo, its components may slowly accrete DM via gravitational capture or scattering, leading to a gradual increase in mass. This mass growth modifies the Keplerian orbital parameters, manifested as a decrease in the semimajor axis and a shortening of the orbital period, thereby accelerating the merger process. In high-density DM environments, this effect can be particularly significant, potentially producing observable phase shifts in the accumulated GW signal, as illustrated in~\cite{arXiv:2106.05043,PRD.96.063001}. In addition, gravitational dynamical friction induced by the DM medium may further enhance the orbital decay rate, introducing additional phase corrections to the GW signal.

Beyond affecting orbital dynamics, DM may alter the internal structure of compact objects, thereby affecting the tidal response in GW waveforms. If DM particles can penetrate neutron star interiors (e.g., with sufficient scattering cross-section or SIDM), their presence modifies the neutron star equation of state (EOS), which in turn changes the mass–radius relation and tidal deformability $\Lambda$. Studies based on relativistic mean-field (RMF) and extended RMF models~\cite{arXiv:2305.02065} indicate that DM admixed neutron stars typically have smaller effective Love numbers compared to pure neutron stars. Consequently, waveforms of binaries in the late inspiral stage last longer with weaker tidal phase contributions, resulting in GW signals that deviate from standard general relativity predictions.

After binary neutron star mergers, DM components may exhibit richer dynamical behavior. If compact cores composed of SIDM exist inside the neutron stars, they may remain gravitationally bound post-merger and orbit around the merger remnant. Models suggest that such DM cores can emit continuous GWs at frequencies of order kilohertz, independent of the normal matter post-merger oscillation modes~\cite{PRD.107.083002}. If such high-frequency signals are detected by future detectors, they could provide compelling evidence for the existence and properties of DM.

For binary black hole systems, scalar fields (e.g., axion or axion-like fields) around rotating black holes may trigger superradiant instabilities, forming self-gravitating “scalar clouds” near the black holes. Studies show that when specific resonance conditions are met, these scalar clouds can drive additional continuous or transient GW emission and back-react on the binary orbital evolution~\cite{arXiv:1907.12089}. Such effects open new avenues for constraining ULDM using GWs.

Several other observationally relevant effects have been explored: additional orbital damping due to DM backgrounds (e.g., enhanced dynamical friction in SIDM environments, which may accelerate orbital decay~\cite{Alonso2024}); mass growth of compact objects due to DM accretion deviating from general relativity predictions, potentially testable through statistical analysis of massive black hole mergers; and weak gravitational lensing effects of DM subhalos along GW propagation paths, causing phase delays or amplitude modulations, representing a new frontier in “GW lensing”~\cite{PRD.112.063055}.

Overall, DM can affect compact binary systems through multiple mechanisms, including accretion, dynamical friction, EOS modification, tidal property changes, post-merger DM core dynamics, and superradiance/resonance effects in black hole binaries. With the improved precision of next-generation GW detectors, such as Cosmic Explorer, Einstein Telescope (ET), and LISA, these effects may become powerful astrophysical probes of DM properties and provide crucial insight into the evolutionary history of compact binaries and predictions for future merger events.

\subsubsection{Effects of Dark Matter on Supermassive Black Hole Mergers and Primordial Black Holes}
SMBHs and their rapid growth have long posed a central challenge in cosmology~\cite{Kormendy2013}. Observations have revealed numerous quasars with masses exceeding $10^9 M_\odot$ at high redshifts ($z>7$), posing significant challenges to standard accretion-based growth scenarios~\cite{Banados2018}. Constrained by the Eddington limit and feedback mechanisms, conventional models struggle to grow seed black holes to observed masses within such a short cosmic time~\cite{Volonteri2010}. Consequently, increasing attention has been given to the possibility of more massive initial seeds, with DM, particularly primordial black holes (PBHs), considered potential key players in SMBH formation and evolution~\cite{Carr_2022,Rubin2001}.

PBHs are thought to form from local collapses triggered by early-universe density fluctuations, phase transitions, or other high-energy processes, with masses spanning from stellar to supermassive scales~\cite{Polnarev1985,Khlopov2008}. Recent cosmological simulations show that in the dense cores of high-redshift DM halos, PBHs can rapidly cluster and merge under dynamical friction, forming massive black hole seeds of $10^{4-5} M_\odot$ within short timescales, providing a natural pathway for the rapid formation of SMBHs~\cite{arXiv:2411.03448}. This mechanism predicts the formation of numerous initial PBH binaries and rapid mergers, generating potentially observable GW signals accessible to future space-based detectors such as LISA.

Beyond PBHs, DM also significantly influences the evolution and mergers of SMBH binaries. A key issue is the so-called “final parsec problem”: after galactic mergers, two black holes may stall at separations of $\sim 1$ pc due to insufficient angular momentum dissipation, preventing further inspiral into the strong gravity regime and efficient GW emission. Studies indicate that self-interacting DM can form dense DM spikes around SMBHs, enhancing dynamical friction and effectively helping binaries overcome the final parsec bottleneck. Gonzalo et al.~\cite{arXiv:2401.14450} further show that velocity-dependent SIDM can not only reliably solve the final parsec problem but also naturally predict the spectral softening observed in low-frequency GW backgrounds in pulsar timing arrays (PTAs), providing theoretically distinguishable signals. Additionally, studies by the Benjamin et al.~\cite{arXiv:2311.18013} suggest that in ULDM environments, short coherence times and rich quasiparticle modes can provide additional long-term angular momentum dissipation channels, also aiding SMBH binaries to traverse the final parsec phase.

The influence of DM on SMBHs is not limited to the merger phase. Regarding accretion and galactic feedback, the distribution and dynamics of DM may profoundly affect black hole growth. If PBHs or dense DM cores exist as seeds, they may trigger early formation of small-scale structures, enabling dense central regions of early galaxies to form more quickly, accelerating gas inflow and driving rapid black hole growth. Moreover, DM-mediated dynamical friction can cause many black holes or stellar remnants to sink rapidly toward galactic centers, enhancing the efficiency of mass assembly in galactic nuclei~\cite{arXiv:2312.04085}. This implies that SMBH formation cannot rely solely on standard gas accretion models but must also account for coupled effects of DM–compact object interactions on central mass assembly.

Despite significant progress, substantial uncertainties remain regarding the true contribution of DM, particularly PBHs, to SMBH growth. For instance, the initial PBH mass function, spatial distribution, and evolution through evaporation or mergers remain poorly constrained. The density structure and dynamical behavior of SIDM or ULDM in actual galaxies also require higher-resolution simulations and observational data. Future GW detectors may directly detect PBH mergers, SMBH binary evolution, and orbit modifications due to background DM, providing new observational tests for theories of DM-driven SMBH growth. In Table~\ref{tabGWsou}, we have summarized the current main findings regarding the influence of DM on GW sources. It is worth noting that although both DM and baryonic environments can cause the orbital energy dissipation and phase shift of GW signals, their physical properties in terms of density distribution and time stability are significantly different. Therefore, while degeneracies exist at the level of orbital phase evolution, their distinct spatial symmetries and temporal behaviors provide promising avenues for disentanglement with high-SNR observations.
\begin{table}[H]
	\caption{Effects of DM on different GW  sources}
	\label{tabGWsou}
	\begin{adjustwidth}{-\extralength}{0cm}
		\begin{tabularx}{\fulllength}{CCCCC}
			\toprule
			\textbf{GW sources} &
			\textbf{DM effect} &
			\textbf{Key DM parameters} &
			\textbf{Affected quantities} &
			\textbf{Detectability} \\
			\midrule
			
			\multirow[m]{3}{*}{EMRIs}
			& Dynamical friction 
			& $\rho_{\rm DM}(r)$, velocity distribution
			& Orbital decay rate, GW phase evolution
			& Phase shift $\Delta\Phi \gtrsim 1$ rad  \\
			
			& Modification of spacetime
			& DM density profile 
			& Orbital precession, eccentricity evolution
			& Cumulative waveform dephasing  \\
			
			& DM accretion onto the secondary
			& $\rho_{\rm DM}$, DM self-interaction strength
			& Secondary mass/spin evolution, GW frequency
			& Subdominant but detectable in high-density or SIDM scenarios \\
			
			\midrule
			
			\multirow[m]{3}{*}{Compact binaries}
			& DM accretion onto compact objects
			& $\rho_{\rm DM}$, scattering cross section
			& Semimajor axis, merger timescale
			& $\Delta t_c/t_c \gtrsim 10^{-3}$  \\
			
			& Dynamical friction 
			& $\rho_{\rm DM}$, velocity dispersion
			& Orbital decay, GW phase accumulation
			& Phase corrections $\Delta\Phi \gtrsim 1$ rad  \\
			
			& DM-modified neutron star EOS
			& DM density, DM--baryon coupling
			& Tidal deformability $\Lambda$, Love number
			& $\mathcal{O}(10\%)$ deviation \\
			
			\midrule
			
			\multirow[m]{3}{*}{SMBH binaries}
			& PBH clustering and seed formation
			& PBH mass function, number density
			& SMBH seed mass, merger rate
			& PBH merger signals  \\
			
			& SIDM-assisted inspiral (final parsec)
			& $\rho_{\rm DM}$, SIDM cross section $\sigma/m$
			& Binary separation, inspiral timescale
			& Low-frequency GW spectrum \\
			
			& ULDM-induced angular momentum dissipation
			& $\rho_{\rm ULDM}$, particle mass $m_{\rm ULDM}$
			& Orbital evolution, GW frequency drift
			& Subtle waveform distortions \\
			
			\bottomrule
		\end{tabularx}
	\end{adjustwidth}
\end{table}
\subsection{Effects of Dark Matter on Gravitational Wave Propagation}\label{DM_pro}
The impact of DM on GW propagation has emerged as an important interdisciplinary topic at the intersection of cosmology and GW physics in recent years~\cite{Oguri2018}. The most direct and widely studied mechanism is lensing amplification and modulation. When GWs propagate through the universe and encounter foreground dense structures, such as PBHs, compact DM clumps, or the centers of high-density DM halos, their wavefronts can be bent, magnified, or interfered with, resulting in observable modulation features~\cite{Takahashi2003}. These lensing effects are physically analogous to electromagnetic wave lensing, but because GWs can traverse nearly all matter distributions, they provide crucial information on the density profiles, clump abundances, and compact DM components at different scales, making GW lensing a unique probe for constraining DM properties. Unlike electromagnetic waves, GWs are insensitive to plasma dispersion, making DM-induced lensing effects cleaner and less affected by baryonic systematics.

Recent studies have explored the feasibility of constraining DM composition using lensed GW signals. Guo et al.~\cite{PRD.106.023018} systematically analyzed the detection capabilities of second-generation, third-generation, and space-based GW detectors for PBH and compact DM lensing effects. Their results indicate that, given sufficiently high signal-to-noise ratios, weak lensing in the wave optics regime can introduce oscillatory modulations in the waveform, allowing future detectors to sensitively constrain the abundance, mass distribution, and clustering properties of small-scale DM structures.

Meanwhile, studies by Oguri et al. and the LIGO Scientific Collaboration~\cite{PRL.122.041103} indicate that if GWs are lensed by compact DM objects with masses in the range $10-10^3 M_\odot$ during propagation, the waveform exhibits interference structures resembling “chirp frequency modulation.” In this scenario, aLIGO could constrain the fraction of compact DM $f_{\text{DM}}$ to a few percent after only one year of observation (example results shown in ~\cite{Jung2019}), providing a novel experimental pathway for probing low-mass PBH DM.

The effects of DM on GW propagation are not limited to lensing. If DM exhibits significant self-interactions, its large-scale behavior may resemble a viscous fluid. In this framework, the effective shear viscosity of DM can introduce additional energy dissipation during GW propagation, causing a mild amplitude attenuation with distance. This mechanism was first proposed by Flauger and Weinberg and has been further quantified in subsequent works~\cite{MNRAS_Letters_502}. Studies show that if the self-scattering cross-section of DM is sufficiently large to manifest fluid-like behavior on galaxy group scales, its shear viscosity can produce observable attenuation for high-frequency GWs. Future third-generation and space-based detectors may be able to detect or constrain this effect, offering a novel experimental test of SIDM models.

Additionally, some studies suggest that if DM exists as a scalar or vector field background , GW propagation may experience time-varying refractive indices, slight shifts in wave speed, or phase distortions. For instance, coherent ULDM fields on galactic scales can induce annual phase modulations in GWs, potentially detectable by LISA~\cite{arXiv:2407.03287}. If dark photons couple to gravity, additional effective medium terms could introduce frequency-dependent group velocities for GWs, which could be tested with future multi-band observations combining LIGO, LISA, and ET~\cite{arXiv:2312.06748}.

Overall, with improving detector sensitivity, GWs provide a powerful avenue for probing DM properties. Lensing effects can test for compact DM components and constrain small-scale structure abundances, while fluid-like attenuation and field-theoretic effects allow examination of self-interactions and the field nature of DM. Future waveform analyses, event statistics, and multi-band observations will offer key insights into the composition and distribution of DM and may ultimately help determine whether it includes PBHs, SIDM, ULDM, or vector DM candidates. Table~\ref{tabpropagation} summarizes the main dark-matter-induced propagation effects on GWs, the associated DM scenarios, and the GW  observables that may be affected in current and future experiments.
\begin{table}[H]
	\caption{Dark matter effects on GW propagation}
	\label{tabpropagation}
	\begin{adjustwidth}{-\extralength}{0cm}
		\begin{tabularx}{\fulllength}{CCCCC}
			\toprule
			\textbf{Propagation effect} 
			& \textbf{DM scenario} 
			& \textbf{Key DM property} 
			& \textbf{Affected GW observables} 
			& \textbf{Detectability} \\
			\midrule
			
			Gravitational lensing 
			& PBHs / compact DM objects 
			& Compactness and abundance 
			& Amplitude magnification, waveform modulation 
			& Possible with high-SNR events \\
			\midrule
			
			Wave-optics interference 
			& PBHs 
			& Lens mass and spatial distribution 
			& Frequency-dependent oscillations, phase distortions 
			& Accessible to ground- and space-based detectors \\
			\midrule
			
			Microlensing by DM substructure 
			& Compact DM clumps / subhalos 
			& Small-scale DM structure 
			& Time-dependent amplitude and phase modulation 
			& Favorable for space-based detectors \\
			\midrule
			
			Viscous attenuation 
			& Self-interacting DM (SIDM) 
			& Effective shear viscosity 
			& Distance-dependent amplitude damping 
			& Potentially testable with future detectors \\
			\midrule
			
			Refractive index modulation 
			& Scalar ULDM 
			& Coherent field background 
			& Periodic phase shifts, propagation speed variation 
			& Particularly relevant for LISA \\
			\midrule
			
			Modified dispersion relation 
			& Vector DM / dark photon background 
			& Effective medium effects 
			& Frequency-dependent phase velocity 
			& Testable via multi-band observations \\
			\bottomrule
		\end{tabularx}
	\end{adjustwidth}
\end{table}

\subsection{Effects of Dark Matter on Gravitational Wave Detectors}\label{DM_det}
The potential impact of DM on GW detectors has become a focal point of both theoretical and experimental research, particularly in scenarios where ULDM couples to SM particles~\cite{Arvanitaki2015}. If such extremely light particles interact with ordinary matter, they can induce periodic mass or mechanical oscillations in test masses within detectors (e.g., interferometer mirrors or beam splitters), leaving observable imprints in the interferometric signals~\cite{Geraci2019}. The long coherence time and narrow-band nature of ULDM-induced signals provide a powerful discriminator against instrumental noise and environmental disturbances. This concept is increasingly recognized as a promising pathway for directly probing DM using GW detectors.

As an example, consider ultralight SFDM~\cite{PRD.100.123512}. Due to its extremely low mass, the corresponding number density is very high, so the field can be treated locally in the detector as a classical scalar background. For simplicity, this scalar field is often approximated as a monochromatic wave:
\begin{align}
	\phi=\phi_{\vec{k}}\cos\left(\omega_k t -\vec{k}\cdot\vec{x}+\theta_{\vec{k}}\right),
\end{align}
and, given its coupling to SM matter (for instance, through modifications of atomic masses or fundamental constants), the action for a test mass (e.g., a mirror or optical element in an interferometer) can be written as:
\begin{align}
	S=-\int m(\phi)\sqrt{-\eta_{\mu\nu}dx^{\mu}dx^{\nu}}.
\end{align}
In the non-relativistic limit, the effective equation of motion can be approximated as:
\begin{align}
	\frac{d^2 x^i}{dt^2}\approx -\kappa \alpha(\phi)\partial_i \phi,
\end{align}
where $\kappa$ represents the gravitational coupling constant, and the DM-matter coupling coefficient $\alpha(\phi)$ is defined by
\begin{align}
	\alpha(\phi)\equiv \frac{d \ln m(\phi)}{d(\kappa \phi)}.
\end{align}
Considering that atomic masses are mainly determined by the QCD scale, $\alpha(\phi)$ can be approximated as
\begin{align}
	\alpha(\phi)\simeq d^*_g \simeq d_g + 0.093(d_{\hat{m}}-d_g),
\end{align}
where $d_g$ represents the coupling of the scalar field to the QCD scale, $d_{\hat{m}}$ represents the coupling of the scalar field back to the average quark mass, and $d^*_g$ represents the effective coupling constant after considering the nuclear structure. By integrating the equation of motion twice, one obtains the oscillatory solution for the test mass position:
\begin{align}
	x^i\simeq d_g^*\kappa \phi_{\vec{k}}\frac{k^i}{m^2_\phi}\sin\left(\omega_k t -\vec{k}\cdot \vec{x}+\theta_{\vec{k}}\right)+\text{const}.
\end{align}
This expression clearly shows that under the influence of a scalar DM field, the test masses (mirrors or other components) in the detector oscillate at the frequency of the background field, producing a continuous periodic interferometric signal.

Based on this mechanism, space-based interferometers such as LISA, Taiji, and TianQin are particularly powerful detection platforms. By employing time-delay interferometry (TDI) techniques, these detectors can effectively suppress laser frequency noise and combine multiple TDI channels to extract potential signals induced by DM~\cite{PRD.100.123512,PRD.108.083007,arXiv:2404.01494}. Previous studies indicate that for certain scalar masses (e.g., $2\times10^{-17}$ eV, $10^{-14}$ eV, and $10^{-12}$ eV), these space detectors could provide constraints on scalar–matter couplings that are significantly tighter than current fifth-force experiments, with enhancement factors reaching tens to hundreds. Some of these results are illustrated in Fig.2 from~\cite{PRD.100.123512}.

For ultralight vector DM, such as dark photons, ground-based GW interferometers can also serve as sensitive probes. Recently, the O3GK run of KAGRA has been used to search for U(1) vector DM~\cite{PRD.110.042001}. This study constrained $\text{U(1)}_\text{B–L}$ dark photons by analyzing the possible coupling-induced forces on test masses composed of different mirror materials. This empirical analysis demonstrates the sensitivity of interferometers to vector DM and provides a roadmap for extending searches to lower masses and longer observation times.

Moreover, theoretical studies have expanded the potential of this approach. For example, a stochastic ultralight scalar DM background could be detected in space interferometers through induced modifications in noise power spectra~\cite{arXiv:2404.01494}. Other works suggest that even in the absence of direct coupling, the gravitational interaction between ULDM and detectors can induce small but cumulative effects via density fluctuations, particularly when the wavelength is comparable to the interferometer arm length~\cite{arXiv:2306.13348}. Notably, the LISA Pathfinder mission was used to attempt the first search for dark photon DM~\cite{arXiv:2301.08736}. Although no significant signal was detected, it provided valuable validation and data analysis experience for future space-based interferometers.

\begin{table}[H]
	\caption{Dark matter effects on GW detectors}
	\label{tab:DM_GW_detectors}
	\begin{adjustwidth}{-\extralength}{0cm}
		\begin{tabularx}{\fulllength}{CCCC}
			\toprule
			\textbf{Detector effect} 
			& \textbf{DM scenario} 
			& \textbf{Key DM property} 
			& \textbf{Affected observables} 
			\\
			\midrule
			
			Test-mass oscillations 
			& Scalar ULDM (SFDM) 
			& Coupling to SM masses 
			& Mirror displacement, phase modulation 
			\\
			\midrule
			
			Periodic interferometric signal 
			& Coherent scalar field 
			& Field coherence and frequency 
			& Continuous narrow-band signals 
			\\
			\midrule
			
			Material-dependent forces 
			& Vector ULDM (dark photons) 
			& Coupling to baryon or lepton number 
			& Differential mirror acceleration 
			\\
			\midrule
			
			Noise spectrum modification 
			& Stochastic ULDM background 
			& Field energy density fluctuations 
			& Excess noise or spectral features 
			\\
			\midrule
			
			Gravitational coupling effects 
			& ULDM without direct SM coupling 
			& Density fluctuations 
			& Weak cumulative displacement effects 
			\\
			\midrule
			
			Pathfinder-scale tests 
			& Vector or scalar ULDM 
			& Effective force sensitivity 
			& Residual acceleration measurements 
			\\
			\bottomrule
		\end{tabularx}
	\end{adjustwidth}
\end{table}
In summary, as the sensitivity of GW detectors continues to improve, using them to detect ULDM via induced oscillations in test masses or interferometric channels is emerging as a new frontier in DM physics. Future challenges primarily lie in distinguishing these signals from environmental noise, such as laser frequency fluctuations, temperature variations, and mechanical resonances, and optimizing data analysis techniques to reduce false positives. By improving noise modeling, enhancing TDI techniques, and conducting longer observation campaigns, this research avenue has the potential to provide a novel window into the nature of ULDM. Table~\ref{tab:DM_GW_detectors} summarizes the main mechanisms through which DM may directly affect GW detectors, the associated DM scenarios, and the interferometric observables that could be exploited in current and future experiments.

\section{Multi-messenger strategies and complementary probes}\label{MultiMess}

Beyond the space-based GW interferometers discussed throughout this review, constraining the nature of DM increasingly relies on multi-messenger approaches that combine GW  observations with complementary probes. Such strategies are particularly valuable for testing DM models across different frequency bands, astrophysical environments, and physical mechanisms, and for reducing degeneracies with standard astrophysical processes.

In this broader context, PTAs provide a powerful and complementary window into DM effects. Operating in the nanohertz frequency range ($\sim10^{-9}$--$10^{-7}$~Hz), PTAs are sensitive to GWs from supermassive black hole binaries and stochastic backgrounds, which may be influenced by DM through mechanisms such as enhanced dynamical friction, modified merger rates, or early primordial black hole clustering. Recent PTA data have therefore opened new opportunities to constrain DM scenarios on galactic and cosmological scales that are not directly accessible to space-based interferometers, as demonstrated by the latest data releases from NANOGrav and EPTA~\cite{NANOGrav2023,EPTA2023}. A comprehensive overview of the astrophysical implications of nanohertz GWs and their relevance to DM physics can be found in Ref.~\cite{Burke-Spolaor2018}.

The combination of PTA observations with space-based interferometer measurements naturally enables a multi-band GW  framework. While PTAs probe the large-scale, low-frequency regime of black hole evolution and stochastic backgrounds, space-based interferometers are sensitive to millihertz signals associated with extreme mass-ratio inspirals, DM spikes in strong gravity, propagation effects, and possible direct detector responses induced by ultralight DM. Joint analyses across these frequency bands can significantly strengthen constraints on DM models and improve discrimination between different physical mechanisms. In particular, PTAs have been shown to be sensitive to coherent oscillatory signals induced by ultralight scalar DM through pulsar timing residuals~\cite{Khmelnitsky2013}, providing a complementary probe to space-based interferometers operating at much higher frequencies.

Electromagnetic observations further enhance this multi-messenger picture by providing independent information on baryonic environments~\cite{Kilpatrick2017}. Measurements of stellar kinematics, gas distributions, and accretion activity in galactic nuclei are essential for constraining astrophysical systematics that may otherwise mimic DM-induced effects in GW  signals. Such observations offer valuable priors for interpreting GW  data and for assessing the robustness of potential DM signatures.

While multi-messenger approaches are often highlighted as a promising avenue for advancing DM studies, a comprehensive review of all relevant probes lies beyond the scope of this work. The primary focus of this review remains on DM effects accessible through space-based GW interferometers. Nevertheless, future progress in this field will likely rely on the combined interpretation of data from space- and ground-based interferometers, pulsar timing arrays, and electromagnetic observations, forming a coherent multi-messenger framework for probing the nature of DM.

\section{Summary and Conclusions}\label{sum}
The evidence for the existence of DM can be considered overwhelming. Contemporary cosmology and structure formation theories require the inclusion of DM to successfully explain observations such as galaxy rotation curves, galaxy cluster dynamics, gravitational lensing, CMB anisotropies, and large-scale structure. Due to its fundamental role in the universe, the intrinsic nature of DM has become a central question at the intersection of modern physics and astrophysics. With the advent of GW detection, the influence of DM on GW sources, propagation, and detectors has gradually emerged as a novel observational avenue, establishing GW astronomy as an increasingly important tool for probing DM.

In this review, we have focused on several widely discussed DM candidates, including WIMPs, various forms of ULDM (covering axions, SFDM, and VFDM such as dark photons), DM halo models that describe the distribution of DM on galactic scales, and SIDM. These candidates differ significantly in terms of their theoretical frameworks, particle mass ranges, couplings to the SM, and astrophysical manifestations, which in turn lead to distinct observational signatures in GW detection.

The effects of DM on GW observations can be summarized from three perspectives: (1) its dynamical influence on GW sources; (2) its impact on GW propagation through the universe; and (3) its influence on the detectors themselves. At the source end, we discussed EMRIs, massive binary black hole mergers, and compact binary systems. In EMRIs, DM affects orbital evolution, energy and angular momentum loss, and the resulting GW waveform via modifications to the gravitational potential around supermassive black holes, formation of dense spikes, and dynamical friction. In massive binary black hole systems, the presence of DM may facilitate rapid black hole growth, enhance merger efficiency, and partially alleviate the “final parsec problem.” For compact binaries, DM accretion can alter stellar masses and orbital parameters, thereby modifying the temporal evolution of GW frequencies and amplitudes.

Regarding propagation effects, DM can influence GW signals through gravitational lensing, affecting waveform shapes, amplitudes, and arrival times. In particular, substructure lensing or dense halo structures may induce interference, phase shifts, or multipath propagation effects, providing potential signals for probing DM distributions. At the detector end, current research mainly focuses on ULDM and its periodic perturbations on interferometer optical paths, test mass motion, or equivalent noise spectra. Such effects can be sought using the high-precision measurement capabilities of current and future GW detectors.

Different classes of GW  signatures probe distinct physical aspects of DM and therefore offer complementary discrimination power among competing scenarios. Effects on GW  sources, such as dynamical friction or density spikes in extreme mass-ratio inspirals and massive black hole binaries, are primarily sensitive to the macroscopic distribution and self-interaction properties of DM, providing a means to distinguish between cuspy and cored profiles or between collisionless and SIDM models. Propagation effects, including gravitational lensing or phase perturbations induced by DM inhomogeneities, are instead more sensitive to the presence of compact or clustered structures, and can therefore probe primordial black holes or other forms of compact DM substructure. In contrast, detector-level effects induced by ultralight DM, such as coherent oscillations of test masses or effective variations of fundamental constants, directly depend on the particle mass and coupling to Standard Model fields, and are largely insensitive to astrophysical environments, offering a particularly clean handle on particle-like DM models.While no single GW  observable is expected to uniquely identify the nature of DM, the diversity of effects accessible to space-based interferometers enables a systematic reduction of the viable model space when different signatures are considered jointly. From a particle-physics perspective, future detections or null results from GW  observations can therefore provide nontrivial constraints on DM masses, interaction strengths, and effective couplings that are complementary to those obtained from laboratory experiments, cosmological observations, and traditional astrophysical probes. In this sense, space-based GW  interferometers are not only discovery instruments, but also precision tools capable of informing and guiding the construction of consistent DM models across a wide range of physical scales.

Finally, the study of DM through GWs naturally benefits from a broader multi-messenger and multi-probe context. Electromagnetic observations, including stellar kinematics, gas dynamics, and accretion signatures in galactic nuclei, provide essential information on baryonic environments and gravitational potentials, helping to disentangle DM effects from astrophysical systematics in GW  signals. Cosmological and large-scale structure observations, such as the cosmic microwave background, galaxy surveys, and weak lensing measurements, place complementary constraints on DM abundance, clustering properties, and interaction strengths across vastly different scales. When combined with GW  data, these probes enable cross-validation of inferred DM properties and improve the robustness of potential interpretations. In this sense, GW  astronomy should be viewed as an integral component of a broader multi-messenger framework, where its unique sensitivity to dynamical and relativistic phenomena provides information that is largely inaccessible to other observational channels.

Looking ahead, the construction and operation of space-based GW detectors will significantly enhance data precision and frequency coverage, making the detection of DM–related effects increasingly feasible. On the theoretical side, more refined DM distribution models, waveform modeling under multi-physics scenarios, and comprehensive multi-messenger observational strategies will jointly improve constraints on DM properties. A deeper understanding of the interactions between DM and GWs will not only advance our knowledge of cosmic composition and structure but may also provide critical insights for exploring physics beyond the SM.
\authorcontributions{
	Conceptualization, Yuezhe Chen, Pan-Pan Wang and Bo Wang; 
	methodology, Yuezhe Chen; 
	formal analysis, Yuezhe Chen; 
	investigation, Yuezhe Chen and Rui Luo; 
	writing---original draft preparation, Yuezhe Chen; 
	writing---review and editing, Yuezhe Chen and Pan-Pan Wang; 
	visualization, Yuezhe Chen; 
	supervision, Pan-Pan Wang, Cheng-Gang Shao; 
	project administration, Pan-Pan Wang; 
	funding acquisition, Pan-Pan Wang. 
	All authors have read and agreed to the published version of the manuscript.
}

\funding{This work was supported by the National Key R\&D Program of China (Grant No.~2024YFC2207500) 
	and the National Natural Science Foundation of China (Grant No.~12405060). }

\institutionalreview{Not applicable.}

\informedconsent{Not applicable.}

\dataavailability{Not applicable. No new data were created or analyzed in this study.}

\acknowledgments{During the preparation of this manuscript, the authors used ChatGPT (GPT-5) for assistance in translating parts of the text from Chinese to English. The authors have reviewed and edited all output and take full responsibility for the content of this publication.}

\conflictsofinterest{The authors declare no conflicts of interest. 
	The funders had no role in the design of the study; in the collection, analyses, or interpretation of data; in the writing of the manuscript; or in the decision to publish the results.} 



\abbreviations{Abbreviations}{
The following abbreviations are used in this manuscript:
\\

\noindent 
\begin{tabular}{@{}ll}
BEC & Bose-Einstein Condensates \\
CDM & Cold Dark Matter \\
CMB & Cosmic Microwave Background \\
DM & Dark Matter \\
EMRI & Extreme Mass Ratio Inspiral \\
ET & Einstein Telescope \\
FDM & Fuzzy Dark Matter \\
GW & Gravitational Wave \\
LISA & Laser Interferometer Space Antenna \\
NFW & Navarro-Frenk-White \\
PBH & Primordial Black Hole \\
PTA & Pulsar Timing Array \\
QCD & Quantum Chromodynamics \\
SFW & Sadeghian-Ferrara-Will \\
SFDM & Scalar Field Dark Matter \\
SIDM & Self-interacting Dark Matter \\
SM & Standard Model \\
SME & Standard Model extensions \\
SUSY & Supersymmetry \\
TDI & Time-delay Interferometry \\
ULDM & Ultralight Dark Matter \\
VFDM & Vector Field Dark Matter \\
SMBH & Supermassive Black Hole \\
WIMP & Weakly Interacting Massive Particles \\

\end{tabular}
}

\appendixtitles{no} 
\appendixstart
\appendix
%
%

\begin{adjustwidth}{-\extralength}{0cm}

\reftitle{References}


 \bibliography{refs.bib}

\PublishersNote{}
\end{adjustwidth}
\end{document}